\documentclass[12pt]{article}
\begin{document}
\title{Analog of Superradiance effect in BEC}
\author{Soumen Basak\\
Physics Department\\   
The Institute of Mathematical Sciences\\
Tharamani\\
Chennai-600113\\
India}
\date{}
\maketitle
\begin{abstract}
  We investigate the scattering of phase oscillation of Bose-Einstein
  Condensate by a 'draining of bathtub' type fluid motion. We derive
  a relation between the reflection and transmission coefficients
  which exhibits existence of analog of 'Superradiance effect' in BEC
  vortex with sink.
  
\end{abstract}
\section{Introduction}
\label{intro}
It has been a hundred years since Einstein's discovery of general
relativity which has been experimentally verified to a high degree of
accuracy. However, on the semiclassical front the absence of direct
experimental observation of the phenomenon of particle emission
predicted by Hawking \cite{hawking} in 1975, still stands out as a
problem. In 1981 though, Unruh \cite{bun,unruh} showed that if a fluid
is barotropic and inviscid, and the flow of the fluid is irrotational,
the equation of motion that fluctuation of the velocity potential of
acoustic disturbance obeys, is identical to that of a minimally
coupled massless scalar field propagating in an effective curved
space-time Lorentzian geometry. One can simulate artificial black hole
\cite{bun,unruh} in this system by trapping the sound wave in a
compact region using radial velocity of the fluid flow. Introduction
of a rotational component \cite{matt,vis,visser} to that radial
velocity will make the artificial black hole an analog of rotating
black hole in physical gravitational system. He also pointed out that
there is possibility of experimental observation of the acoustic
analog of Hawking radiation from regions of flow of inviscid and
barotropic fluids behaving as outer trapped surfaces ('acoustic event
horizons'). This model is sufficiently rich to enable probing of
almost all the kinematic aspects of general relativity but not its
dynamics, because dynamics of a fluid system is governed by Euler
equation and continuity equation but dynamics of a gravitational
system is governed by Einstein's equations.

In our earlier work \cite{soumen, majumdar}, using the analogy between
a shrinking fluid vortex ('drain bathtub') modelled as a (2+1)
dimensional fluid flow with a sink at the origin and a rotating (2+1)
dimensional black hole with an ergosphere, it has been shown that a
scalar sound wave is reflected from such a vortex with an
amplification for a specific range of frequencies of the incident
wave, depending on the angular velocity of rotation of the vortex.
This is analogous to superradiant scattering \cite{zeld, star, dew}
which occurs in the case of rotating black holes in asymptotically
flat spacetime. A more detailed analysis this phenomenon in the low
frequency range has also been done in the next paper \cite{ majumdar}.
The reflection coefficient exhibit its frequency dependence which is
consistent with the results in \cite{ majumdar}. We have also
discussed the possibility of observation of this phenomenon especially
for inviscid fluids like He-II, where vortices with quantized angular
momentum may occur.

Although we have demonstrated the analog of superradiance in the
superfluid helium, BEC is a far cleaner system which can be
manipulated to produce the kind of metric we have (DB) in a much
easier manner. That is why BEC seems to be better suited for this
purpose. Recent experimental investigation of Bose-Einstein
condensation \cite{kurn,mewes,anderson} has further improved the
prospects of experimental realization of the analog of kinematical
effects of gravity.

In this paper we investigate the possibility of the acoustic analog of
superradiance scattering from hydrodynamic vortex with a sink in
Bose-Einstein condensate. Linear acoustic perturbations in such a
system are shown to scatter from the ergoregion with an enhancement in
amplitude, for a restricted range of frequencies of the incoming wave.

The paper is organised follows. First, in section(2), we briefly
discuss how effective Lorentzian geometry emerges from the propagation
of phase oscillation in BEC. In section(3) we briefly describe the
properties of effective geometry that emerges from propagation of
phase oscillations through BEC. In section(4) we demonstrate analog
of superradiance effect in BEC for position dependent background
configuration. We conclude with some remarks in section(5).

\section{Effective Geometry in BEC}
A Bose-Einstein condensate \cite{legg,smith} is a collection of
indistinguishable bosons such that all of them are in same
single-particle quantum state $\phi(\vec{r})$. If there is no
interaction between the bosons, then the second quantized Hamiltonian
for N such bosons is given by,

\small
\begin{equation}
H = \sum_{i=1}^{N} \left(\frac{p^{2}_{i}}{2 m} + V(\vec{r}_{i})\right)
\end{equation} 

and the corresponding  wave function of this system is,

\small
\begin{equation} 
\Phi(\vec{r}_{1},\vec{r}_{2}\,.\,.\,.\,.\,.\vec{r}_{N})=\phi({\vec{r}_{1}})
\,\,\phi({\vec{r}_{2}}),\,.\,.\,.\,.\,\,.\,.,\phi({\vec{r}_{N}})
\label{wavfun}
\end{equation} 

Even there are very week interaction between the atoms, at ${0}^{o}k$
almost all the atoms in the condensate remains in the same
single-particle state. If the number of boson is large enough then the
time evolution of such a condensate is determined by Gross-Pitaevskii
equation,

\small
\begin{equation}
i\hbar\partial_t\Phi=\left(-\frac{\hbar^2}{2m}\nabla^2+ V_{\rm
    ext}+\frac{4\pi a\hbar^2}{m}|\Phi|^2\right)\Phi ,
\end{equation} 

where $m$ is the mass of the atoms, $a$ is the scattering length, and
we normalise to the total number of atoms $\int d^3{\mathbf{x}}
|\Phi({\mathbf{x}},t)|^2=N$.\\

Expressing $~\Phi(\vec{r},t)~$ in terms of its amplitude
$~\sqrt{\rho(\vec{r},t})~$ and phase$~\Psi(\vec{r},t)~$, the above
wave equation(\ref{wavfun}) is converted into two first order partial
differential equation, 

\small
\begin{eqnarray}
\partial_{t}{\Psi}&=&-\frac{1}{2m}\,(\vec{\nabla}\Psi.\vec{\nabla}\Psi)
- V  -\frac{4\pi a\hbar}{m}\rho + \frac{\hbar^{2}}{2m}\frac{1}{\sqrt{\rho}}
\nabla^{2}\sqrt{\rho}\nonumber\\
\nonumber\\
\partial_{t}{\rho}&=&-\frac{1}{m}(\vec{\nabla}\rho.\vec{\nabla}\Psi + 
\rho \nabla^{2}\Psi)
\end{eqnarray}

Now linearising $\rho$ and $\Psi$ around there background values
$\rho_{0}$ and $\Psi_{0}$ respectively, 

\small
\begin{eqnarray}
\partial_{t}{\Psi_{1}}&=&-\frac{1}{m}\vec{\nabla}\Psi_{0}
.\vec{\nabla}\Psi_{1}- \frac{4\pi a\hbar}{m}\rho + 
\frac{\hbar^{2}}{2m}\left\{\frac{1}{2\sqrt{\rho_{0}}}
\nabla^{2}\left(\frac{\rho_{1}}{\sqrt{\rho_{0}}}\right) 
- \rho_{1}\frac{\nabla^{2}\sqrt{\rho_{0}}}{2\,\rho^{\frac{3}{2}}}) 
\right\}\nonumber\\
\nonumber
\partial_{t}{\rho_{1}}&=&-\frac{1}{m}\,\vec{\nabla}(\rho_{0}
\vec{\nabla}\Psi_{1})-\frac{1}{m}\,\vec{\nabla}(\rho_{1}
\vec{\nabla}\Psi_{0})
\label{pertequ}
\end{eqnarray}

Here $\rho_{1}$ and $\Psi_{1}$ are the perturbed values of $\rho$ and
$\Psi$ respectively.

If we neglect the quantum pressure term, then the combination of above
equations(\ref{pertequ}) results the following covariant differential
equation, 

\small
\begin{equation}
\frac{1}{\sqrt{-g}}\,\partial_\mu(\sqrt{-g}g^{\mu\nu}\partial_\nu\Psi)=0
\label{kgequ}
\end{equation}

where,

\small
\begin{equation}
g^{\mu\nu}(t,\vec r) = 
{1\over \rho_0 c}
\left( \matrix{-1&\vdots&-v_0^j\cr
               \cdots\cdots&\cdot&\cdots\cdots\cdots\cdots\cr
               -v_0^i&\vdots&(c_s^2 \delta^{ij} - v_0^i v_0^j )\cr } 
\right).               
\end{equation}
\\

The speed of sound$(c)$ and the background velocity$(v_0)$ are defined
as,

\begin{equation}
c(r)=\frac{\hbar}{m}\sqrt{4\pi a \rho_{0}(r)}\hspace{0.5in}v_0^i=\frac{1}{m}
\partial_{i}\Psi_{0}
\label{sndvel}
\end{equation}

Therefore the propagation of the phase oscillations in Bose-Einstein
condensate is same as the propagation of a minimally coupled massless
scalar field in an effective Lorentzian geometry which is determined
by the background velocity and background density of the condensate
and also by the speed of sound in the condensate \cite{garay,anglin,
  barcelo}.

Since in this paper, we investigate the possibility of the acoustic
analog of superradiance (a phenomenon that we call `superresonance'),
i.e., the amplification of a sound wave by reflection from the
ergo-region of a rotating acoustic black hole, we choose the
so-called `draining bathtub' type of fluid flow \cite{vis}, which is
basically a (2+1) dimensional flow with a sink at the origin. A two
surface in this flow, on which the fluid velocity is everywhere
pointing towards the sink, and the radial velocity component exceeds
the local sound velocity everywhere, behaves as an outer trapped
surface in this `acoustic' spacetime, and is identified with the
(future) event horizon of the black hole analog. Thus, the velocity
potential for the flow has the form \cite{vis} (in polar coordinates
on the plane),
\small
\begin{equation}
\psi(r,\phi)~=~A ~\log r~+~B~\phi ~, 
\label{vpot}
\end{equation}

where, $A$ and $B$ are real constants. This leads to the velocity
profile,

\small
\begin{equation}
\vec{v}=-\,{\frac{A}{r}}\hat{r}+{\frac{B}{r}}\hat{\phi} ~. 
\label{vel}
\end{equation}

For the velocity potential given in equation(\ref{vpot}), the analog black
hole metric is (2+1) dimensional with Lorentzian signature, and is given
by,

\begin{equation}
{ds}^2=\left(\frac{\rho_{0}}{c}\right)^{2}\left[-\left(c^2-\frac{A^2+B^2}
{r^2}\right){dt}^2-\frac{2A}{r}\,dr\,dt-2B\,d\phi\,dt+{dr}^2+r^2\,
{d\phi}^2 \right]
\end{equation}

where $c$ is the velocity of sound. In our earlier work we had
neglected the conformal factor in the effective metric because the
back ground density and hence the background pressure of fluid were
assumed to be constant. In this paper we have relaxed this condition
by considering position dependence of the density profile, which leads
to the position dependence of the speed of sound also. 

Recently Slatyer and Savage \cite{savage} have suggested the following density
profile \cite{smith} which could be realized(at least nearly) in
future BEC experiments, 

\small
\begin{equation}
\rho_{0}(r)=\rho_{\infty}\,\,\frac{(r-r_{0})^{2}}{(r-r_{0})^{2} 
+ 2\,\sigma^{2}}
\end{equation}

where $\sigma$ is the spatial scale of variance, the distance over
which the wave function of the Bose-Einstein Condensate tends to it
bulk values, when the wave function is subjected to a localized
perturbation. For analytical simplicity we have considered $r_{0} = 0$
case in this paper. The sound speed(\ref{sndvel}) corresponding to
this density profile is,

\begin{equation}
c(r)=c_{\infty}\,\,\sqrt{\frac{r^{2}}{r^{2} 
+ 2\,\sigma^{2}}}
\end{equation}

where,
\begin{equation}
c_{\infty}= \frac{\hbar}{m}\sqrt{4\pi a\rho_{\infty}}
\end{equation}

Considering the above velocity profile(\ref{vel}) and new density
profile yields new locations of analog of event horizon and
ergosphere. As for the Kerr black hole \cite{wald} in general
relativity, the radius of the ergosphere is given by the vanishing of
$g_{00}$, i.e.,at

\small
\begin{equation}
r_e =
\left[\frac{(A^{2} + B^{2})} {2\,\,c_{\infty}^{2}}\,\left\{1\,
+\,\left(1\,+\frac{\,8\,c_{\infty}^{2}\,\sigma^{2}}{A^{2}+ B^{
2}}\right)^\frac{1}{2}\right\}\right]^{\frac{1}{2}}
\end{equation}

The metric has a (coordinate) singularity at, 

\small
\begin{equation}
r_h = \left[\frac{A^{2}} {2\,\,c_{\infty}^{2}}\,\left\{1\,+\,\left
(1\,+\frac{\,8\,c_{\infty}^{2}\,\sigma^{2}}{A^{2}}\right)^\frac{1
}{2}\right\}\right]^{\frac{1}{2}}
\end{equation}

, which signifies the horizon, i.e.,the boundary of the outer trapped
surface.

It should be noticed that both the radius of the horizon and the
ergosphere are larger than what we found in our earlier work assuming
constant density profile. This result is expected because in this case
the speed of sound is decreasing towards the centre of the vortex and
hence it becomes lesser than fluid velocity at larger value of radial
coordinate.

\section{Superresonance Scattering}

From the components of the draining vortex metric it is clear that the
(2+1)-dimensional curved spacetime possesses isometries that
correspond to time translations and rotations on the plane. The
solution of the massless Klein Gordon equation(\ref{kgequ}) can therefore be
written as,
\begin{equation}
\Psi(t,r,\phi)=Re\{R(r)~e^{-i~\omega~t}~e^{i~m~\phi}\}
\end{equation}
where $~\omega~$and$~m~$ are real and positive. In order to make
$~\Psi(t,r,\phi)~$ single valued , m should take  integer values.\\
Then the radial function $~R(r)~$ satisfies,

\small
\begin{equation}
\frac{d^{2}R(r)}{d{r}^{2}} + P_{1}(r)~\frac{dR(r)}{dr}
+Q_{1}(r)~R(r)=0~,
\label{radcom}
\end{equation}

where,

\small
\begin{eqnarray}
P_{1}(r)=\frac{r^2 + 2\,{\sigma }^2}{c_{\infty}\,r\,
\left\{r^4\,\left( r_{h}^2 + 2\,{\sigma }^2 \right) -r_{h}^4\,
\left( r^2 + 2\,{\sigma }^2 \right) \right\} }\,P_{2}(r)\nonumber
\end{eqnarray}

\small
\begin{eqnarray}
P_{2}(r)=c_{\infty}\,r_{h}^4 - 2\,i 
\,W\,r^2\,r_{h}^2\,{\sqrt{r_{h}^2 + 2\,{\sigma }^2}} + 
\frac{c_{\infty}\,r^4\,\left( r_{h}^2 + 2\,{\sigma }^2 \right) \,
\left( r^2 + 6\,{\sigma }^2 \right) }{{\left( r^2 + 2\,{\sigma }
^2 \right) }^2}\nonumber
\end{eqnarray}

\small
\begin{eqnarray}
Q_{1}(r)=-\left( \frac{{\sqrt{r_{h}^2 + 2\,{\sigma }^2}}}
{c_{\infty}^2\,r^2\,\left( r^2 - r_{h}^2 \right) \,
\left( r^2\,r_{h}^2 + 2\,\left( r^2 + r_{h}^2 \right) \,
{\sigma }^2 \right) } \right) Q_{2}(r) \nonumber
\end{eqnarray}

\small
\begin{eqnarray}
Q_{2}(r)= 2\,i\,B\,c_{\infty}\,m\,r_{h}^2\,\left( r^2 + 
2\,{\sigma }^2 \right)  + r^4\,{\sqrt{r_{h}^2 + 2\,{\sigma }^2}}\,
\left\{ {c_{\infty}}^2\,m^2 - W^2\,\left( r^2 + 2\,{\sigma }^2 
\right)  \right\} \nonumber
\end{eqnarray}

and,

\small
\begin{equation}
W=\omega - \frac{B\,m}{r^{2}}
\end{equation} 

Now in order to eliminate the imaginary part from the $~P_{1}(r) ~$ and
$~Q_{1}(r)~$ we made the following substitution, 
\small
\begin{equation}
R(r)=e^{\,i\,\omega\,f_{1}(r)} \,e^{ -\,i\,m\,f_{2}(r)}\,\,G(r)
\end{equation} 

\small
\begin{eqnarray}
f_{1}(r)=\frac{{\sqrt{r_{h}^2 + 2\,{\sigma }^2}}}{2\,c_{\infty}\,
\left( r_{h}^4 + 6\,r_{h}^2\,{\sigma }^2 + 8\,{\sigma }^4 \right) }
\,\,f_{3}(r)\nonumber
\end{eqnarray}

\small
\begin{eqnarray}
f_{2}(r)=\frac{B}{2\,c_{\infty} \,r_{h}^2\,\left( r_{h}^2 + 4\,
{\sigma }^2 \right) }\,f_{4}(r)\nonumber
\end{eqnarray}

\small
\begin{eqnarray}
f_{3}(r)={\left( r_{h}^2 + 2\,{\sigma }^2 \right) }^2\,\log\left 
(r^2 - r_{h}^2\right) - 4\,{\sigma }^4\,\log \left\{r^2\,r_{h}^2 
+ 2\,\left( r^2 + r_{h}^2 \right) \,{\sigma }^2\right\}\nonumber
\end{eqnarray}

\small
\begin{eqnarray}
f_{4}(r)=\left( r_{h}^2 + 2\,{\sigma }^2 \right) \,\log 
\left(\frac{r^2 - r_{h}^2}{r^2}\right) + 2\,{\sigma }^2\,
\log \left(\frac{r^2\,r_{h}^2 + 2\,\left( r^2 + r_{h}^2 \right) 
\,{\sigma }^2}{r^2}\right)\nonumber
\end{eqnarray}

In terms of new radial function $G(r)$ the radial
equation(\ref{radcom}) takes the following form, \small
\begin{equation}
\frac{d^{2}G(r)}{d{r}^{2}} + S_{1}(r)~\frac{dG(r)}{dr}
+S_{2}(r)~G(r)=0
\end{equation}
\label{radreal}

where,

\small
\begin{eqnarray}
S_{1}(r)=\frac{3\,r^4 + r_{h}^4}{r\,(r^4 - r_{h}^4)} 
- \frac{2\,r}{r^2 + 2\,{\sigma }^2} + \frac{2\,r\,r_{h}^4}
{\left( r^2 + r_{h}^2 \right) \,\left( r^2\,r_{h}^2 + 
2\,\left( r^2 + r_{h}^2 \right) \,{\sigma }^2 \right) }\nonumber
\end{eqnarray}

\small
\begin{eqnarray}
S_{2}(r)=\frac{r^2\,\left( r_{h}^2 + 2\,{\sigma }^2 \right) }
{c_{\infty}^2\,{\left( r^{2} - {r_{h}}^{2} \right) }^2\,{\left( r^2\,r_{h}^2 
+ 2\,r^2\,{\sigma }^2 + 2\,r_{h}^2\,{\sigma }^2 \right) }^2}
\,S_{3}(r)\nonumber
\end{eqnarray}

\small
\begin{eqnarray}
S_{3}(r)=W^2\,r^4\,\left( r^2 + 2\,{\sigma }^2 \right) \,
\left( r_{h}^2 + 2\,{\sigma }^2 \right)  + c_{\infty}^2\,m^2\,
\left\{r_{h}^4\,(r^2 + 2\,{\sigma }^2) - r^4\,
\left( r_{h}^2 + 2\,{\sigma }^2 \right)  \right\}\nonumber
\end{eqnarray}
 
Now we introduce tortoise coordinate $~r^{*}~$ through the
equation,

\small
\begin{equation}
\frac{d}{dr^{*}}=\left(1-\frac{A^{2}}{r^{2}~c^{2}}\right)\frac{d}{dr}
\end{equation}

which implies that,

\small
\begin{eqnarray}
r^{*}=r - \frac{2\,r_{h}\,{\sigma }^3 }{{\sqrt{\frac{r_{h}^2}{2} 
+ {\sigma }^2}}\,\left( r_{h}^2 + 4\,{\sigma }^2 \right) } 
\,\cot^{-1}\left(\frac{\sqrt{2}\,r_{h}\,\sigma } {r\,{\sqrt{r_{h}^2
+ 2\,{\sigma }^2}}} \right) - \frac{r_{h}\,\left( r_{h}^2 + 2\,
{\sigma }^2 \right) \,\tanh^{-1} (\frac{r}{r_{h}})}
{r_{h}^2 + 4\,{\sigma }^2}\nonumber
\end{eqnarray}

This tortoise coordinate spans the entire real line as opposed to r
which spans only the half-line. The horizon at $r=r_{h}$ maps
to$~ r^{*}\rightarrow -\infty~$, while $~
r\rightarrow \infty~$corresponds to$~ r^{*}\rightarrow
+\infty~$. 

Let us now define a new radial function$~G(r)$ as,

\small
\begin{equation}
G(r)=\frac{{\sqrt{r^2 + 2\,{\sigma }^2}}}{r^{\frac{3}{2}}}\,H(r)
\end{equation}

Substituting this in equation(\ref{radreal}) we observe that $~H(r)~$
satisfies the differential equation, 

\small
\begin{equation}
P(r)\,\frac{d^{2}H(r)}{d{r}^{2}} + Q(r)\,\frac{dH(r)}{dr}
+ V(r)\,H(r)=0
\end{equation}

where,

\small
\begin{eqnarray}
P(r)=\left\{1 -\,\frac{ r_{h}^4\,\left( r^2 + 2\,{\sigma }^2 \right)}
{r^4\,\left( r_{h}^2 + 2\,{\sigma }^2 \right) } \right\}^{2}\nonumber
\end{eqnarray}

\small
\begin{eqnarray}
Q(r)=\frac{2\,r_{h}^4\,\left( r^2 + 4\,{\sigma }^2 \right) }
{r^5\,\left( r_{h}^2 + 2\,{\sigma }^2 \right) }\left\{1 -\,
\frac{ r_{h}^4\,\left( r^2 + 2\,{\sigma }^2 \right)}{r^4\,
\left( r_{h}^2 + 2\,{\sigma }^2 \right) } \right\}\nonumber
\end{eqnarray}

\small
\begin{eqnarray}
V(r)=\frac{\left( r^2 + 2\,{\sigma }^2 \right) }{c_{\infty}^2\,r^2} \,\, W^2
+ \frac{\left( r^{2} - r_{h}^{2} \right) \,\left( r^2\,r_{h}^2 + 2\,r^2\,
{\sigma }^2 + 2\,r_{h}^2\,{\sigma }^2 \right) }{{\left( r^2 + 2\,{\sigma }^2 \right) }^2\,{\left( r_{h}^2 + 2\,{\sigma }^2 \right) }^2}\,\, S(r)  
\nonumber
\end{eqnarray}

and,
 
\small
\begin{eqnarray}
S(r)&=&-\frac{51\,r_{h}^4\,{\sigma }^4}{r^8} - \frac{42\,
r_{h}^4\,{\sigma }^6}{r^{10}} - \frac{{\sigma }^2\,
\left\{ 35\,r_{h}^4 + 2\,{\sigma }^2\left( 3 + 4\,m^2 \right) 
\,\left( r_{h}^2 + 2\,{\sigma }^2 \right)  \right\} }{2\,r^6}\nonumber\\
\nonumber\\
&+&\frac{\left( 1 - 4\,m^2 \right) \,\left( r_{h}^2 + 2\,
{\sigma }^2 \right) }{4\,r^2} - \frac{5\,r_{h}^4 -\,{\sigma }^2 
\,\left( 5 - 4\,m^2 \right) \,\left( r_{h}^2 + 2\,
{\sigma }^2 \right) }{4 r^4}\nonumber
\end{eqnarray}

Now in terms of tortoise coordinate one obtains the modified
differential equation as,

\small
\begin{equation}
\frac{d^{2}H(r^{*})}{d{r^{*}}^{2}} + V(r)H(r^{*})=0
\label{radequ}
\end{equation}

We analyse this differential equation in two distinct radial regions,
viz., near the sonic horizon, i.e., at $r^{*}\rightarrow -\infty$ and
at asymptopia, i.e., at $~ r^{*}\rightarrow +\infty~$.

Our choice of coordinates and substitutions are with a view to
transform differential equation(\ref{radcom}) into a
form(\ref{radequ}) which is simple enough to identify ingoing and
outgoing modes both in asymptotic and near horizon regions. In
addition we have managed to eliminate the first derivative term from
the differential equation which leads to constancy of Wronskian of the
differential equation. This will facilitate the calculations of the
transmission and reflection coefficients. Since the second term of $
V(r)$ in equation(\ref{radequ})tends to zero both in asymptotic region
and near the horizon, the first term in $V(r)$ is the only important
term.

In the asymptotic region, the above differential equation(\ref{radequ})
can be written approximately as,

\begin{equation}
\frac{d^{2}H(r^{*})}{d{r^{*}}^{2}}+
\frac{\omega^{2}}{c^{2}_{\infty}}~H(r^{*})=0
\label{radasym}
\end{equation}

This can be solved trivially,

\begin{equation}
H(r^{*})={\cal R}_{\omega m}~\exp~\left(i~\frac{\omega}
{c_{\infty}}~r^{*}\right)+\exp~\left(-i~\frac{\omega}
{c_{\infty}}~r^{*}\right)
\label{solasym}
\end{equation}

The first term in equation(\ref{radasym}) corresponds to reflected wave
and the second term to the incident wave, so that R is the reflection
coefficient in the sense of potential scattering. It is not difficult
to calculate the Wronskian of the solutions(\ref{solasym}); one
obtains,

\begin{equation}
{\cal W}(+\infty)~=\,2~i~\frac{\omega}{c_{\infty}}~
\left(1-|{\cal R}_{\omega m}|^2\right)~.
\end{equation}

Similarly , near the horizon the above differential equation can be
written approximately as,

\small
\begin{equation}
\frac{d^{2}H(r^{*})}{d{r^{*}}^{2}}+\left[\frac{\left( r_{h}^2 + 2\,
{\sigma }^2 \right) }{r_{h}^2} \,\frac{(\omega-m~\Omega_{H})^{2}}
{c^{2}_{\infty}}\right]\,H(r^{*})=0
\end{equation}

where$~\Omega_{H}~$ is the angular velocity of the sonic horizon. We
impose the physical boundary condition that of the two solutions of
this equation, only the ingoing one is physical, so that one has,

\small
\begin{equation}
H(r^{*})={\cal T}_{\omega m}~\exp~\left\{-i\,\,\sqrt{\frac{\left( r_{h}^2 + 2\,
{\sigma }^2 \right) }{r_{h}^2}}\,\,\frac{(\omega-m~\Omega_{H})} 
{c_{\infty}}~r^{*}\right\}
\label{solhor}
\end{equation}

Once again, it is easy to calculate the Wronskian of this solution;
one obtains,

\begin{equation}
{\cal W}(-\infty)~=\,2~i~\left(\frac{\omega~-~m\Omega_H}{c_{h}}
\right)~|{\cal T}_{\omega m}|^2~.
\end{equation}

Now using these approximate solutions (\ref{solasym},\ref{solhor}) of the
above differential equation together with their complex conjugates and
recalling the fact two linearly independent solutions of this
differential equation(\ref{radequ}) must lead to a constant Wronskian,
it is easy to show that,

\small
\begin{equation}
1-|{\cal R}_{\omega m}|^{2}=\left(1-\frac{m~\Omega_{H}}{\omega}\right)\,
\frac{c_{\infty}}{c_{h}}\,\,|{\cal T}_{\omega m}|^{2}
\label{relation}
\end{equation}

where,

\small
\begin{equation}
c_{h}=\left(\frac{r_{h}^2}{r_{h}^2 + 2\,{\sigma }^2 }\right)^{\frac{1}{2}}
\,{c_{\infty}}
\end{equation}
\\

Here$~{\cal R}_{\omega m}~$ and $~{\cal T}_{\omega m}~$ are the
amplitudes of the reflection and transmission coefficients of the
scattered wave, respectively. It is obvious from
equation(\ref{relation}) that, for frequencies in the range
$~0~<~\omega~<~m~\Omega_{H}$, the reflection coefficient has a
magnitude larger than unity. This is similar to the amplification
relation that emerges in our earlier analysis of superresonance
\cite{soumen,majumdar}. The only difference here is the appearance of
a factor $~c_{\infty}\big{/}c_{h}~$. This is due to the fact that
speed of sound in Bose-Einstein condensate depends on radial
coordinate.

\section{Conclusion}

In this paper we have shown that when acoustic disturbances propagate
through a Bose-Einstein condensate, they get scattered by the flow of
the condensate and in the particular frequency range acoustic
disturbances are reflected with an amplitude greater than the
amplitude of the incident wave. The frequency range is determined by
angular momentum of the incident wave and angular velocity of the
horizon. The extra energy that reflected waves carry is taken from the
rotational energy of the condensate. As a result vortex motion is
slowed down and eventually stops when all the rotational energy of the
condensate is extracted out of it. At other frequencies acoustic
disturbances transmit through the condensate and ultimately get
trapped inside the Dumb hole. This is analogous to the superradiance
effect that occurs exclusively in the case of rotating black holes in
physical gravitational systems. Because of the discreteness of the
parameter B of rotational component of the velocity profile
(\ref{vpot})(as evident in BEC), the angular momentum of the acoustic
black hole will also be proportional to an integer and hence causes
the energy flux to change discretely making observation of the
phenomenon easier than that in actual black holes. This also manifests
itself in the reflection coefficient whose spectrum has equally spaced
peaks of varying strengths all of which are multiples of the minimum
strength.  These interesting aspects are worth being studied
quantitatively in greater detail.

Unlike our earlier work we have considered, following
the suggestion of Slatyer and Savage \cite{savage}, a particular type
of spatial dependence of the background density of the condensate.
Presence of spatial scale of variation of the condensate wave function
has nontrivial effect on the artificial Lorentzian geometry that
emerges from BEC. Due to spatial dependence of the background density
of the fluid, speed of sound in such a system varies when it
propagates though the fluid. This spatial dependence modifies the
relation between reflection and transmission coefficients of the
scattered waves by a factor $c_{\infty}/c_{H}$, but does not affect
the cutoff frequency of the superradiance effect.  In the limit
$\sigma$ tends to zero the background density and hence the speed of
sound in the fluid become constant.  As a check we note that in this
limit we recover the results of our earlier works .

\begin{center}
  ACKNOWLEDGEMENT\\
  I would like to thank my advisor Prof. Parthasarathi Majumdar for
  several useful discussions. I am also grateful to Saha Institute of
  Nuclear Physics for their local hospitality, where a part of this
  work was carried out.
\end{center}

\end{document}